\documentclass[12pt]{article}
\begin{document}
\baselineskip=16pt

\title
{\begin{Huge}
Weighing the Black Hole via\\ Quasi-local Energy\\
\end{Huge}
\vspace{0.3in} }
\author{Yuan K. Ha\\ Department of Physics, Temple University\\
 Philadelphia, Pennsylvania 19122 U.S.A. \\
 yuanha@temple.edu \\    \vspace{.1in}  }
\date{Published 10 July 2017}
\maketitle
\vspace{.2in}
\begin{abstract}
\noindent
We set to weigh the black holes at their event horizons in various spacetimes and obtain 
masses which are substantially higher than their asymptotic values. In each case, the
horizon mass of a Schwarzschild, Reissner-Nordstr{\"o}m, or Kerr black hole is found to be
twice the irreducible mass observed at infinity. The irreducible mass does not contain
electrostatic or rotational energy, leading to the inescapable conclusion that particles
with electric charges and spins cannot exist inside a black hole. This is proposed as the
External Energy Paradigm. A higher mass at the event horizon and its neighborhood
is obligatory for the release of gravitational waves in binary black hole merging. We
describe how these mass values are obtained in the quasi-local energy approach and 
applied to the black holes of the first gravitational waves GW150914.\\

\end{abstract}

\newpage
\vspace*{.5in}
\noindent
{\bf 1. \hspace{0in} Black Hole Theorems}\\

\noindent
Black holes are natural outcomes of solutions to Einstein's equation. Since the discovery
of the first gravitational waves from binary black hole merging in 2015 [1], black holes
are now real astrophysical bodies. They are as legitimate as the elementary particles whose
existence is confirmed indirectly. They may be abundant in the Universe and their properties
can be investigated from the gravitational waves emitted in binary black hole merging.

\indent
   A number of important theorems on black holes have been established between 1965 and 2005. 
They provide the conceptual framework and predict the properties of classical black holes
in terms of temperature, entropy, irreversibility, thermodynamics, as well as energy conditions.
They are known as\\

\noindent
(1) Singularity Theorem (1965) [2],\\
(2) Area Non-decrease Theorem (1972) [3],\\
(3) Uniqueness Theorem (1975) [4],\\
(4) Positive Energy Theorem (1979) [5],\\
(5) Horizon Mass Theorem (2005) [6].\\

\noindent
The first four theorems listed above have been extensively discussed in general relativity
for many years and we take for granted that their contents are well known to general relativists.
It is the last theorem, the Horizon Mass Theorem, which we shall discuss in this report and apply it
to the black holes of the first gravitational waves GW150914.\\

\vspace{.2in}
\noindent
{\bf 2. \hspace{0in} Quasi-local Energy}\\

\noindent
The Horizon Mass Theorem is the final outcome of the quasi-local energy approach [7] applied to 
black holes. The quasi-local energy gives the total energy within a spatially bounded region instead of
defining locally the energy density for a gravitational system. It is obtained from a Hamiltonian-Jacobi 
analysis of the Hilbert action in general relativity and it is uniquely suited for investigating the dynamics
of the gravitational field [8]. The mass of a black hole can be found anywhere by calculating the total energy
contained in a Gaussian surface enclosing the black hole at a given coordinate distance. The usual mass of a 
black hole is the asymptotic mass seen by an observer at infinity.\\
\indent
   A black hole has the strongest gravitational potential energy of any gravitational system. This energy exists
outside the black hole. An observer at a distance sees the total of the constituent mass contained at the horizon
and the intermediary potential energy. Since gravitational potential energy is always negative and extends 
throughout all space, the closer an observer gets to the black hole, the less gravitational potential energy
the observer will see. Thus, the mass of the black hole increases as the observer gets near the horizon. This is
a unique situation for black holes since for any other physical object, the gravitational potential energy is far
insignificant compared to its mass and therefore the mass appears to be the same at all distances of observation.

\indent
   The Brown and York expression for quasi-local energy is given in terms of the total mean curvature
of a surface bounding a volume for a gravitational system in four-dimensional spacetime.
It is given in the form of an integral
\begin{equation}
E = \frac{c^{4}}{8\pi G} \int_{^{2}B} d^{2}x \sqrt{\sigma} (k-k^{0}) ,
\end{equation}
where $\sigma$ is the determinant of the metric defined on the two-dimensional surface $^{2}B$ ;
$k$ is the trace of extrinsic curvature of the surface and $k^{0}$ , the trace of curvature of a
reference space. For asymptotically flat reference spacetime, $k^{0}$ is zero.\\

\vspace{.2in}
\noindent
{\bf 3. \hspace{0in} Horizon Mass Theorem}\\

\noindent
The Horizon Mass Theorem can be stated as follows,\\ 

\newpage
\noindent
{\bf Theorem.} {\em For all black holes; neutral, charged or rotating, the horizon mass is always
twice the irreducible mass observed at infinity.}\\

In notation, it has the simple form,
\begin{equation}
M_{\rm horizon} = 2M_{irr} ,
\end{equation}
where $M_{irr}$ is the irreducible mass. The derivation of this theorem is given fully in Ref. 6. 
The Horizon Mass Theorem relates the mass of a black hole at the event horizon to its irreducible mass.
It is an exact result obtained only with the knowledge of the spacetime metrics of Schwarzschild, 
Reissner-Nordstr{\"o}m, and Kerr without further assumption. It is a new addition to the previous theorems
on classical black holes.\\
\indent
In order to understand the Horizon Mass Theorem, it is necessary to introduce the various mass terms involved.

\begin{enumerate}
  \item The {\em asymptotic mass} is the mass of a neutral, charged or rotating black hole
        including electrostatic and rotatinal energy. It is the mass observed at infinity
        used in the various spacetime metrics.
  \item The {\em horizon mass} is the mass which cannot escape from the horizon of a neutral,
        charged or rotating black hole. It is the mass of the black hole observed at the horizon.
  \item The {\em irreducible mass} is the final mass of a charged or rotating black hole when
        its charge or angular momentum is removed by adding external particles to the 
        black hole. It is the mass observed at infinity.
\end{enumerate}

\noindent
The Horizon Mass Theorem is remarkable in that the mass contained at the event horizon depends only on the
irreducible mass of the black hole. The irreducible mass does not contain electrostatic or rotational energy.
This leads to the surprising conclusion that the electrostatic and rotational energy exist only outside the
black hole. They are all external quantities. An asymptotic observer investigating a charged or rotating
black hole is in fact exploring a Schwarzschild black hole with external energies in between.\\

\vspace{.2in}
\noindent
{\bf 4. \hspace{0in} External Energy Paradigm}\\

\noindent
There are profound implications which follow from the Horizon Mass Theorem. Since all electric field lines
terminate at electric charges and electrostatic energy is external, this indicates that electrical particles
cannot exist inside a black hole. They can only stay at the surface. Similarly, since rotational energy is external,
any particle with angular momentum also cannot exist inside the black hole and must stay outside, as required by the
Horizon Mass Theorem. Together, this implies that all elementary particles possessing charges and spins can only stay
outside the horizon. If a black hole is formed from the collapse of a star made of ordinary matter, the result will be
a hollow and thin spherical shell with all constituent mass at the horizon. This is a radical view of the black hole
and it follows inescapably from the property of the irreducible mass. We are thus led to introduce a new paradigm for
black holes to be called the External Energy Paradigm.  {\em All energies of a black hole are external quantities. 
Matter particles are forbidden inside a black hole and can only stay outside or at the horizon}. These energies include
constituent mass, gravitational energy, heat energy, electrostatic energy and rotational energy. It explains naturally
why the entropy of a black hole is proportional to the area and not to the volume because matter particles are all at
the surface.\\

\vspace{.2in}
\noindent
{\bf 5. \hspace{0in} Schwarzschild Black Hole}\\

\noindent
The total energy contained in a sphere enclosing the black hole at a coordinate distance $r$
is given by the expression [6,7,9]\\
\begin{equation}
E(r) = \frac{rc^{4}}{G} \left[ 1- \sqrt{ 1- \frac{2GM}{rc^{2}} } \right],
\end{equation}
where $M$ is the mass of the black hole observed at infinity, $c$ is the speed of light and $G$ is
the gravitational constant. At the horizon, the Schwarzschild radius is $r = R_{S} = 2GM/c^{2}$ .
Evaluating the expression in Eq.(3), we find that the metric coefficient 
$g_{00}=(1-2GM/rc^{2})^{1/2}$ vanishes identically and the horizon energy is therefore
\begin{equation}
E(r)=\left( \frac{2GM}{c^{2}} \right) \frac{c^{4}}{G} = 2Mc^{2} .
\end{equation}
The horizon mass of the Schwarzschild black hole is simply twice the asymptotic mass
$M$ observed at infinity. The negative gravitational energy outside the black hole is as
great as the asymptotic mass.

\indent
Equation (3) can be used to evaluate the mass seen by an observer at any distance $r$. We show some particular
values in Table 1.\\

\vspace{.1in}
\begin{table}[h]
\caption{Mass of black hole observed at a distance $r$.}
\label{tab:1}
\begin{center}
\begin{tabular}{cc}
\hline
Coordinate $r$ in $R_{S}$  &  Mass in $M_{\infty} = M$  \\ 
\hline\\ 
1 & 2.000 \\
2 & 1.172 \\
3 & 1.101 \\
4 & 1.072 \\
5 & 1.056 \\
6 & 1.046 \\
7 & 1.039 \\
8 & 1.033 \\
9 & 1.029 \\
10 & 1.026 \\
100 & 1.003 \\
$\infty$ & 1.000 \\  
\\
\hline

\end{tabular}
\end{center}
\end{table}

\vspace{.1in}
From the listed values, it is seen that $90\%$ of the negative potential energy lies within a distance of two
Schwarzschild radii outside the horizon, i.e. $R_{S} < r < 3R_{S}$. At a distance of $r = 100 R_{S}$, the mass
is seen to be only $0.3\%$ higher than the asymptotic value $M$. An observer at that location is approaching a 
near flat spacetime.\\

\vspace{.2in}
\noindent
{\bf 6. \hspace{0in} Reissner-Nordstr{\"o}m Black Hole}\\

\noindent
We investigate next the Reissner-Nordstr{\"o}m black hole in the quasi-local energy approach. The total energy 
of a charged black hole contained within a radius at coordinate $r$ is now given by [6]
\begin{equation}
E(r) = \frac{rc^{4}}{G} \left [ 1 - \sqrt{ 1 - \frac{2GM}{rc^{2}} + \frac{G Q^{2}}{r^{2}c^{4}} } \right ] .
\end{equation}
Here, $M$ is the mass of the black hole including electrostatic energy observed at infinity
and $Q$ is the electric charge. At the horizon radius
\begin{equation}
r_{+} = \frac{GM}{c^{2}} + \frac{GM}{c^{2}} \sqrt{1 - \frac{Q^{2}}{GM^{2}} } ,
\end{equation}
the metric coefficient $g_{00}$ given by the square root in Eq.(5) also vanishes and the horizon energy becomes
\begin{equation}
E(r_{+}) = \frac{r_{+}c^{4}}{G} = Mc^{2} + Mc^{2} \sqrt{ 1 - \frac{Q^{2}}{GM^{2}} }.
\end{equation}
For the Reissner-Nordstr{\"o}m black hole, the irreducible mass which is obtained when the charge is removed by 
adding oppositely charged particles has the expression
\begin{equation}
M_{irr} = \frac{M}{2} + \frac{M}{2} \sqrt{ 1 - \frac{Q^{2}}{GM^{2}} }.
\end{equation}
Combining Eqs.(7) and (8), we find the horizon energy to be exactly twice the irreducible energy
\begin{equation}
E(r_{+}) = 2M_{irr}c^{2} ,
\end{equation}
which depends only on the mass of the black hole when the charge is neutralized.\\

\vspace{.2in}
\noindent
{\bf 7. \hspace{0in} Kerr Black Hole}\\

\noindent
The rotating black hole is considerably more complicated to handle in the quasi-local energy approach because
one is comparing a rotating spacetinme with a fixed spacetime. It is therefore not possible to give an exact 
analytical expression as in the previous two cases. An approximate energy expression [10] is available for a 
slowly rotating black hole with angular momentum $J$ and angular momentum parameter $\alpha = J/Mc$ , where 
$0 < \alpha \ll GM/c^{2}$ ,
\begin{eqnarray}
E(r) & = & \frac{rc^{4}}{G} \left[ 1 - \sqrt{ 1 - \frac{2GM}{rc^{2}} + \frac{\alpha^{2}}{r^{2}} } \right ] \nonumber \\
     &   & + \frac{\alpha^{2}c^{4}}{6rG} \left [ 2 + \frac{2GM}{rc^{2}} 
        + \left ( 1 + \frac{2GM}{rc^{2}} \right ) \sqrt{ 1 - \frac{2GM}{rc^{2}} + \frac{\alpha^{2}}{r^{2}} } \right ] + \cdots    \end{eqnarray}
Again, with the horizon radius of the Kerr black hole 
\begin{equation}
r_{+} = \frac{GM}{c^{2}} + \sqrt { \frac{G^{2}M^{2}}{c^{4}} - \frac{J^{2}}{M^{2}c^{2}} }
\end{equation}
and the definition of the irreducible mass
\begin{equation}
M_{irr}^{2}  = \frac{M^{2}}{2} + \frac{M^{2}}{2} \sqrt { 1 - \frac{J^{2}c^{2}}{G^{2}M^{4}} } ,
\end{equation}
we arrive at a very good approximate relation for the horizon energy
\begin{equation}
E(r_{+}) \simeq 2 M_{irr}c^{2} + O(\alpha^{2}) . 
\end{equation}

\indent
    For general and fast rotations, the energy can be accurately obtained by numerical evaluation in the 
teleparallel equivalent of general relativity [11]. The result shows almost perfectly that the horizon mass
is twice the irreducible mass. For an exact and impeccable relationship, we have to employ a formula known
for the area of a Kerr black hole valid for all rotations [12], i.e.
\begin{equation}
A = 4 \pi ( r_{+}^{2} + \alpha^{2} ) = \frac {16 \pi G^{2}M_{irr}^{2}}{c^{4}} .
\end{equation}
This area is exactly the same as that of a Schwarzschild black hole with an asymptotic mass $M_{irr}$ ,
\begin{equation}
A = 4 \pi R_{S}^{2} = 4 \pi \left( \frac{2GM_{irr}}{c^{2}} \right)^{2} .
\end{equation}
As shown earlier, the horizon mass of such a Schwarzschild black hole is twice the irreducible mass. 
We have therefore established the Horizon Mass Theorem for all black hole cases. The profound consequence
of this theorem is that elementary particles with charges or spins cannot exist inside a black hole.\\ 

\vspace{.2in}
\noindent
{\bf 8. \hspace{0in} Black Holes of GW150914}\\

The discovery of gravitational waves GW150914 by LIGO confirmed the existence of two black holes in a binary
system. They merged to form a single black hole with the release of gravitational energy. We realize that the
energy of the gravitational waves comes from outside the black holes and not from their interiors. The waves are
generated predominately from near the horizon and they are gravitationally redshifted as they propagate to infinity.
The horizon energy therefore becomes important. Without a higher mass at the event horizon and its neighborhood,
there can be no gravitational waves emitted in black hole merging.\\
\indent
  The two black holes of GW150914 are rotating black holes. For a Kerr black hole, rotation necessarily contributes to
the overall mass observed at infinity. To find the irreducible mass of the Kerr black hole, we need to know the 
dimensionless spin parameter $a$, which is the ratio of the angular momentum $J$ to the maximum possible angular
momentum, i.e.
\begin{equation}
a = \frac{J}{ \left( \frac{GM^{2}}{c} \right) } = \frac{Jc}{GM^{2}} .
\end{equation}
The irreducible mass, from Eq.(12), is then given by
\begin{equation}
M_{irr} = \left [ \frac{M^{2}}{2} + \frac{M^{2}}{2} \sqrt{ 1 - a^{2} } \right]^{\frac{1}{2}} 
\end{equation}
and the horizon mass can be found as twice the irreducible mass,
\begin{equation}
M(r_{+}) = 2M_{irr} = \left[ 2M^{2} + 2M^{2} \sqrt{ 1 - a^{2} } \right]^{ \frac{1}{2} }. \\
\end{equation}

\indent
  For the black holes of GW150914 [13], the primary black hole has a mass of $36 M_{Sun}$ and an average model
spin parameter $a = 0.32$. The secondary black hole has a mass of $29 M_{Sun}$ and average model spin parameter
$a = 0.44$. The final black hole has a mass of $62 M_{Sun}$ and a spin parameter $a = 0.67$ . Accordingly,
$3 M_{Sun}$ of energy is released as gravitational waves to infinity as report by LIGO, i.e.
\begin{equation}
36 M_{Sun} + 29 M_{Sun} = 62 M_{Sun} + 3 M_{Sun} .
\end{equation}
\noindent
However, this $3 M_{Sun}$ wave energy has been significantly reshifted and Eq.(19) does not account for the 
missing energy. Without additional source of energy, the waves cannot propagate away from the deep potential of 
the black hole. To understand the energy of the waves at the source, we need to know the mass of the black holes
at the event horizon. For the primary black hole, the horizon mass is found to be $71 M_{Sun}$ and for the secondary
black hole, a horizon mass of $57 M_{Sun}$ . The final black hole has a horizon mass of $116 M_{Sun}.$  

\noindent
In an ideal merging, the energy at the horizon would follow the equation
\begin{equation}
71 M_{Sun} + 57 M_{Sun} = 116 M_{Sun} + ( 3 M_{Sun} + 3 M_{Sun} ) + 6 M_{Sun} .
\end{equation}

\noindent
In this account, the energy for the redshift is now available. Analysis of a mass removed from the surface of
a black hole shows that the energy required has the same magnitude as the energy of the waves observed at infinity [14].
The total energy required to release $3 M_{Sun}$ of wave energy to infinity is therefore $3 M_{Sun} + 3 M_{Sun} = 6 M_{Sun}$.
The remaining $6 M_{Sun}$ of the energy is for uncertainties in LIGO data. These mass values are additional properties
for the binary black hole merger of GW150914.\\
\indent
   We may further provide the rotational energy of the black holes by comparing the asymptotic mass with the irreducible
mass. The rotational mass of the primary black hole is $36 M_{Sun} - 35.5 M_{Sun} = 0.5 M_{Sun}$ , while that of the
secondary black hole is $29 M_{Sun} - 28.25 M_{Sun} = 0.75 M_{Sun}$ . The final black hole has a higher rotational mass
that is $62 M_{Sun} - 58 M_{Sun} = 4 M_{Sun}.$  The initial black holes in a binary system generally have different
spin orientations. Thus most of the rotational energy of the final black hole comes from the orbiting energy of the
binary system.\\

\newpage
\noindent
{\bf 9. \hspace{0in} Conclusion}\\

\noindent
The detection of the first gravitational waves GW150914 shows the need to understand the energy of the black hole at
the event horizon. This was first emphasized by the author in 2003 in the paper ``The Gravitational Energy of a Black Hole"
[14] before the success of LIGO was certain. At the end, there was the remark:
\begin{quotation} 
\noindent
   {\em Therefore, in detecting any gravitational signals from a black hole collision such as that proposed in the 
   LIGO project, any conclusion about the strength of the signals near its source should be based on the black hole
   energy formula.} 
\end{quotation} 
\noindent
The quasi-local energy approach and the Horizon Mass Theorem are indispensable tools in the latest development of
general relativity.\\

\end{document}